\def\be{\begin{equation}}
\def\ee{\end{equation}}
\def\bea{\begin{eqnarray}}
\def\eea{\end{eqnarray}}
\def\bse{\begin{subequations}}
\def\ese{\end{subequations}}

\documentclass[prl,twocolumn,amsmath,amssymb,superscriptaddress,showpacs]{revtex4}

\usepackage{graphicx}

\usepackage{dcolumn}

\usepackage{bm}

\begin{document}

\title{Proposal to stabilize and detect half-quantum vortices
in strontium ruthenate thin films: Non-Abelian braiding
statistics of vortex matter in a ${p_x}+i{p_y}$ superconductor.}
\author{Sankar Das Sarma}
\affiliation{Condensed Matter Theory Center, Department of Physics, University of Maryland, College Park, MD 20742}
\author{Chetan Nayak}
\affiliation{Microsoft Research, Project Q, Kohn Hall, University of California,
Santa Barbara, CA 93108}
\affiliation{Department of Physics and Astronomy, University of California,
Los Angeles, CA 90095-1547}
\author{Sumanta Tewari}
\affiliation{Condensed Matter Theory Center, Department of Physics, University of Maryland, College Park, MD 20742}

\date{\today}

\begin{abstract}
We propose a simple way to stabilize half-quantum vortices in
superconducting strontium ruthenate, assuming the order parameter is
of chiral $p_x + ip_y$ symmetry, as is suggested by recent
experiments. The method, first given by Salomaa and Volovik in the
context of Helium-3, is very naturally suited for strontium
ruthenate, which has a layered, quasi-two-dimensional, perovskite
crystal structure. We propose possible
experiments to detect their non abelian-braiding statistics. These
experiments are of potential importance for topological
quantum computation.
\end{abstract}

\pacs{74.20.Rp, 03.67.Lx, 03.67.Mn, 03.67.Pp}

\maketitle

\paragraph{Introduction.}
The realizable prospect for topological quantum computation
using fractional quantum Hall states in two-dimensional ($2D$)
semiconductor structures has been discussed recently in the literature
\cite{DasSarma05,Stern05,Bonderson05,Day,Bonesteel05}.
In particular, non-Abelian braiding statistics, manipulated by
braiding particles as desired, is the robust quantum mechanical
resource underlying quantum computation in these systems.
Such non-Abelian quasiparticle statistics arise naturally in certain
exotic classes of $2D$ fractional quantum Hall states (e.g. the
so-called Pfaffian quantum Hall state which may exist at
the $\sigma_{xy}=\frac{5}{2} \frac{e^2}{h}$ plateau
\cite{DasSarma05,Stern05,Bonderson05,Day}), giving rise
to the recent excitement about the possibility of quasiparticle
manipulation and braiding (and eventually to topological
quantum computation) in high-mobility semiconductor
heterostructures. Given the great interest in non-Abelian
statistics and topological quantum computation, it
is relevant to ask whether other physical systems could be
studied experimentally to investigate non-Abelian braiding
statistics. In this Letter, we suggest quantum
vortex matter in superconducting strontium ruthenate
as a prospective candidate for non-Abelian statistics
and topological quantum computation. The physical
basis for this proposal is the possible ${p_x}+i{p_y}$ symmetry
of the superconducting order parameter in strontium
ruthenate. This order parameter symmetry admits
half-quantum vortices, which are analogous to
the fractionally-charged non-Abelian quasiparticles
of the Pfaffian fractional quantum Hall state.

\paragraph{Superconductivity in Sr$_2$RuO$_4$.}
Spin triplet ($\vec{S}=1$), odd parity ($\vec{L}=1$) superfluidity
is realized in Helium-3 \cite{Vollhardt}. Recently, a body of
evidence, including sensitivity to impurity \cite{Mackenzie},
nuclear magnetic resonance (NMR) Knight shift \cite{Ishida}, and
polarized neutron scattering \cite{Duffy} have strongly indicated
that spin-triplet superconductivity is also realized in strontium
ruthenate (Sr$_2$RuO$_4$) \cite{Rice-Physicstoday}.
Furthermore, within the superconducting phase of this metallic
oxide, time-reversal symmetry breaking has been confirmed by
muon spin rotation \cite{Luke}. The existence of a two-component
order parameter is suggested by small angle neutron scattering
exploring the flux lattice in the mixed state \cite{Kealey}. Taken
together, these experiments point to the conclusion
\cite{Rice-Physicstoday,Mackenzie-Review}
that strontium ruthenate, with a highest $T_c$
of about $1.5$ K, realizes the simplest unitary $T$-breaking
$p$-wave state allowed in its tetragonal crystal structure,
namely $p_x+ip_y$, which is
analogous to the A-phase of He 3. Very recently, phase
sensitive superconducting quantum interference device (SQUID)
measurements \cite{Nelson-Science} have given almost conclusive
evidence of $p$-wave superconductivity \cite{Rice-Science}
in Sr$_2$RuO$_4$.

For spin triplet superconductors, the order parameter is a matrix
in spin space \cite{Vollhardt},
\begin{equation}
\Delta_{\alpha\beta}(\bm{k}) =
\sum_{\mu=1}^3d_\mu(\bm{k})\left(\sigma_\mu i
\sigma_2\right)_{\alpha\beta}.\label{orderparametermatrix}
\end{equation}
 Here $\alpha$, $\beta$ are spin
indices, ${\bm k}$ is the wave vector, $\sigma_{1,2,3}$ are the
Pauli matrices, and ${\bm d}({\bm k})$ is a complex 3-vector.
Orbital p-wave symmetry implies $d_{\mu}({\bm k}) = \sum_{j=1}^3
d_{\mu j}\,{\hat k}_j.$ The tensor field $d_{\mu j}$ is the
appropriate order parameter.
For He 3 A-phase, or $p_x+ip_y$ superconductor,
\begin{equation}
d_{\mu j}=\Delta_0 (T)\hat{d_{\mu}}(\hat{m}_j+i\hat{n}_j)e^{i\phi}
\label{orderparameter},
\end{equation} where $\Delta_0(T)$ is the temperature-dependent magnitude
of the order parameter, $\hat{m}$ and $\hat{n}$ are mutually
orthogonal unit vectors in orbital space, and $\phi$ is the phase
angle. Physically, $\hat{d}$ is the unit vector in spin space on
which the projection of the Cooper pair spin is zero, and
$\hat{l}=\hat{m}\times\hat{n}$ is a preferred direction in the
orbital space giving the direction of the Cooper pair angular
momentum.

For topological classification of vortices \cite{Salomaa-Review},
note that the full symmetry group G=SO(3)$\times$ SO(3)$\times$
U(1), implying rotational symmetries in the spin and the orbital
spaces, and the global gauge symmetry. The $p_x+ip_y$ state has
residual symmetry H=U(1)$\times$ U(1)$\times$ Z$_2$, where the
first U(1) is for spin rotation symmetry about the vector
$\hat{d}$, the second U(1) is the combined gauge-orbital symmetry,
and the Z$_2$, most important for our purposes, signifies a
discrete  combined gauge-spin rotation symmetry
\cite{Volovik-Mineev} of the order parameter,
$(\hat{d}, \phi)\rightarrow(-\hat{d},\phi+\pi).$ The first
homotopy group of the degeneracy space R then corresponds to
$\pi_{1}$(R)=$\pi_{1}$(G/H)=
$\pi_{1}$((S$^{2}$$\times$SO$_3$)/Z$_2$)=Z$_4$, the cyclic group
with 4 elements. Here S$^2$ denotes the surface of a sphere. The
vortices, then, correspond to only four classes, with vorticities
$N=-\frac{1}{2}, 0, \frac{1}{2}, 1=-1$. The surprising fact is the
existence of $N=\pm \frac{1}{2}$, which correspond to half flux
quantum per vortex.
The half-quantum vortex is crucial for non-Abelian statistics in this
system.

\paragraph{Stabilizing Half-Quantum Vortices.}
The simplest representation of the class $N=\pm\frac{1}{2}$ is
given by,
\begin{eqnarray}
&&\hspace*{-0.5in}\hat{d}=\hat{x}\cos(\theta/2)+\hat{y}\sin(\theta/2),\hspace{0.1in}
\phi=\pm\theta/2, \nonumber\\
&&\hspace*{-0.5in}\hat{l}={\rm{const}}.
\label{representation}\end{eqnarray} Here, it has been assumed,
without any loss of generality, that $\hat{d}$ lies in the $x-y$
plane and $\theta$ is the polar angle with the vortex core as the
origin. The flux enclosed by a half-vortex is half the flux
quantum, $\frac{1}{2}\phi_0$, where $\phi_0=\frac{hc}{2e}$. The
existence of the half-vortex is possible because $\hat{d}$ can
change sign upon circulating the core, and the phase angle $\phi$
can simultaneously change by $\pi$ to keep the order parameter,
Eq.~\ref{orderparameter}, unchanged.

However, in real systems, half-vortices are energetically costlier
than vortices with $N=1, 2$ ( a vortex with $N=2$ is continuously
deformable to $N=0$). This is because of small, but non-zero,
spin-orbit or dipole energy,
$E_{\rm{so}}=-\Omega_{\rm{so}}(\hat{d}.\hat{l})^2$, which binds
$\hat{l}$ and $\hat{d}$ parallel, and so penalizes spatial
modulation of $\hat{d}$ around the vortex core. The texture for a
half-quantum vortex requires $\hat{d}$ and $\hat{l}$ to be
nonparallel over a region of macroscopic size. Because of
$E_{\rm{so}}$, this region forms a domain wall terminated by the
vortex, expending energy proportional to the volume of the domain
wall. Another way of looking at this is to note that in the case
of dipole-locking of $\hat{d}$ and $\hat{l}$, the extra Z$_2$
symmetry of H is absent. The first homotopy group of R is then
$\pi_{1}$(S$^2$$\times$SO$_3$)=Z$_2$. There are only two classes
of vortices, the uniform class, $N=0$ (a vortex with $N=2$ remains
continuously deformable to this class), and the class with
vorticity $N=1$. Half-quantum vortices are no longer possible.
Hence, to stabilize the half-quantum vortices, the spin-orbit
energy must be neutralized.

A simple way to neutralize the spin-orbit energy, first proposed
in the context of He 3 \cite{Salomaa-Volovik}, is very naturally
suited to strontium ruthenate. In He 3 A-phase, one has to use an
artificial parallel plate geometry so that the direction of the
Cooper pair angular momentum, $\hat{l}$, is constrained to be
along the plate normals, $\hat{n}$. In the presence of a
sufficiently strong magnetic field $\hat{H}\parallel \hat{n}$, the
Cooper pair spins will align themselves with $\hat{H}$. The vector
$\hat{d}$ is then in the plane of the parallel plates, expending
the maximum spin-orbit energy. However, this energy cost is a
constant for {\textit {all}} types of vortices. $\hat{d}$ is also
free to rotate in the plane to $-\hat{d}$ spending no additional
spin-orbit energy. In this way one can avoid the large domain wall
energy cost for the half-quantum vortices making them competitive
with other topological excitations.

However, this restrictive geometry is completely natural for
strontium ruthenate. It is a metallic oxide with highly
two-dimensional (2D), layered, tetragonal crystal structure. In
fact, the crystal structure of Sr$_2$RuO$_4$ is almost identical
to that of perovskite (La, Sr)$_2$CuO$_4$ high temperature
superconductor \cite{Rice-Physicstoday}. Two dimensional,
correlated RuO$_2$ layers are separated by intermediate strontium
reservoir. Electronic conduction is mediated by $d$-orbitals of
ruthenium ions in 2D planes. This highly planar structure is also
evident by the superconducting state properties such as the upper
critical field $H_{c2}$, which is about 20 times smaller when
applied parallel to the $c$-axis than when applied parallel to the
$ab$ plane.

 Sr$_2$RuO$_4$ thus naturally realizes a $p_x+ip_y$ superconductor
 in a  parallel plate geometry. It is believed
that $\hat{l}$ is in the $c$-direction, and that $\hat{d}$ is
pinned to $\hat{l}$ due to spin-orbit energy, although
conclusive proof is lacking \cite{Mackenzie-Review}. If this is
indeed the case, the most direct way to stabilize half-quantum
vortices is to simply apply a magnetic field, strong enough to
overcome the spin-orbit energy, parallel to the $c$-axis. In the
case of He 3, Ref.~\cite{Salomaa-Volovik} claimed that at low
enough temperatures $T\ll T_c$, vortices with $N=\frac{1}{2}$ are
energetically slightly more favorable than those with $N=1, 2$. In
the absence of a definitive proof of even the symmetry of the
order parameter being $(p_x + ip_y)$-type and lack of knowledge of
other system parameters, such microscopic energy comparisons for
Sr$_2$RuO$_4$ are premature. However, half-quantum vortices are
expected to be at least competitive in energy and may even be more
favorable in the presence of magnetic fields simply because of
smaller vorticity.

The required magnetic field $H$ can be estimated from the
spin-orbit or dipole coupling constant $\Omega_{\rm {so}}$.
Unfortunately, we are unaware of any reliable theoretical estimate
of this quantity. In He 3 A-phase, the required $H$ was estimated
to be greater than $\sim 25$ G \cite{Salomaa-Volovik}. In
Sr$_2$RuO$_4$, a recent experimental estimate of the spin-orbit
decoupling field is $\sim 250$ G \cite{Murakawa}. We suggest
applying a magnetic field of this order of magnitude in order
to decouple $d$ from $l$. Note that this can be done without
destroying superconductivity since the upper critical field in the
$c$-direction is still larger: $H_{c2\parallel c}\sim750$ G
\cite{Mackenzie-Review}.

\paragraph{Non-Abelian Statistics of Half-Quantum Vortices.}
Once stabilized, the half-quantum vortices may be detectable by
exploiting their statistics under interchange of positions. In the
following, we recapitulate the arguments leading to their
non-abelian braiding statistics \cite{Read00,Ivanov,Stern04}.
An interpretation
of the half-quantum vortex can be given, when the subsystems of up
and down spins are uncoupled, in terms of a single-quantum vortex
for only one of the spins and zero vorticity for the other spin
component \cite{Leggett, Ivanov}. This can be seen by explicitly
writing out the components of the order parameter matrix,
Eq.~\ref{orderparametermatrix}. Ignoring, for the moment, the
orbital dependencies of all quantities, we get,
$\Delta_{\uparrow\uparrow}=\Delta_0\exp(i\phi)(-d_x+id_y),
\hspace{.1in}
\Delta_{\downarrow\downarrow}=\Delta_0\exp(i\phi)(d_x+id_y)$ and
$\Delta_{\uparrow\downarrow}=\Delta_{\downarrow\uparrow}=\Delta_0\exp(i\phi)d_z$.
Using the form of $\hat{d}$ and $\phi$ in
Eq.~\ref{representation}, with $\phi=\theta/2$ (positive
half-quantum vortex), it is easy to see that,
$\Delta_{\uparrow\uparrow}=-\Delta_0,
\hspace{0.1in}\Delta_{\downarrow\downarrow}=\Delta_0\exp(i\theta)$
and $\Delta_{\uparrow\downarrow}=\Delta_{\downarrow\uparrow}=0$.
Because the phase-angle $\phi=\theta$ changes by $2\pi$ upon
circulating the core, the down spin pairs see a full single
quantum vortex, while the up spin pairs see no vortex at all.
Because of this property, the Bogoliubov-de-Gennes equations
separate into two pieces: the part for the down spin component is
identical to that for an ordinary single-quantum vortex in a
$p$-wave superconductor \cite{Kopnin}, albeit for spinless
electrons, while the part for the up spin component is devoid of
any vortex. As in Ref.~\cite{Kopnin}, the low-energy
spectrum in the vortex core for the down spins is given by
$E_n=n\omega_0$, where $\omega_0=\frac{\Delta_0^2}{\epsilon_F}$
with $\epsilon_F$ the Fermi energy, and $n$ an integer. For $n=0$,
there is a zero energy state in the vortex core.

It is interesting to note that the Bogoliubov quasiparticles
corresponding to the levels $E_n$ are linear superpositions of
particle and hole operators from the {\it same} species, namely,
down spin electrons, $\gamma_i^{\dagger}=u\psi_{i
\downarrow}^{\dagger}+v\psi_{i \downarrow}$, where $\psi_{i
\downarrow}$ is the electron annihilation field for down spins at
the location of the  $i$-th vortex. They satisfy
$\gamma_{i}^{\dagger}(E_n)=\gamma_{i}(-E_n)$. The zero energy
state, then, is a Majorana fermion state, satisfying
$\gamma_{i}^{\dagger}(0)=\gamma_i(0)$. Each half-quantum vortex,
$i$, has a zero energy Majorana fermion $\gamma_i$ in its core. In
the case of $2m$ such spatially separated vortices, the $2m$
Majorana operators can be pairwise combined to produce $m$ complex
fermion operators, each of which can be occupied or unoccupied in
the ground state. The ground state, thus, is $2^{m}$ fold
degenerate. Further, using the property that $\gamma_i$'s carry
odd charge with respect to the gauge field of a single-quantum
vortex, that is, $\gamma_i\rightarrow-\gamma_i$ for a phase change
of $2\pi$, it follows that \cite{Ivanov}, upon interchange of two
neighboring half-quantum vortices $\gamma_i$, $\gamma_{i+1}$,
\begin{eqnarray}
\label{eqn:statistics}
\gamma_i &\rightarrow& \gamma_{i+1},\nonumber\\
\gamma_{i+1}&\rightarrow& -\gamma_i\nonumber\\
\gamma_j&\rightarrow&\gamma_j\hspace{0.3in}j\neq i, j\neq i+1
\end{eqnarray}

\paragraph{Observing non-Abelian Braiding Statistics.} 
These topological properties are identical to
those of the Pfaffian quantum Hall state, which
is a candidate description of the observed fractional
quantum Hall plateau with $\sigma_{xy}=\frac{5}{2}\,\frac{e^2}{h}$
in two-dimensional electron gases \cite{DasSarma05,Stern05,Bonderson05,Day}.
This state is distinguished by the non-abelian braiding statistics
of its quasiparticles, which are precisely (\ref{eqn:statistics}).
This state has recently been the subject of much interest
as a platform for fault-tolerant quantum computation
\cite{DasSarma05}. Recent papers \cite{Stern05,Bonderson05},
elaborating on an earlier proposed experiment
\cite{Fradkin98} explain how non-Abelian statistics
could be directly detected through quantum interference
measurements. In this paper, we suggest that
the same underlying physics could arise in Sr$_2$RuO$_4$.
Following ref. \onlinecite{DasSarma05}, we note
that the two possible states of a pair of vortices in a $2D$ ${p_x}+i{p_y}$
superconductor (corresponding to the presence or absence of a neutral
fermion) can be used as the two states of
a qubit. When the two vortices are far apart, no
local measurement can distinguish the two states
of the qubit, so they form a decoherence-free subspace.
However, there are important differences
between Sr$_2$RuO$_4$ and 2DEGs which lead
to practical differences, which we discuss below.

Vortices in Sr$_2$RuO$_4$ are neutral, unlike
quasiparticles in a quantum Hall liquid.
Therefore, a direct quantum interference measurement
of quasiparticle transport as proposed in \cite{DasSarma05}
would not apply to Sr$_2$RuO$_4$.
However, the Josephson relation tells us that
the motion of a vortex across a line connecting
points $1$ and $2$ causes a phase slip,
thereby inducing a voltage drop between $1$ and $2$
according to
\begin{equation}
\label{eqn:Josephson}
{V_x} = \frac{h}{e}\,j_y^{\rm vortex}
\end{equation}
When vortices become depinned,
either as a result of non-equilibrium effects or thermal
fluctuations, there is dissipation.

A second important difference is that, even in
a thin film, there are many $2D$ layers in
a Sr$_2$RuO$_4$ crystal. Hence, a vortex which
runs through $N$ layers actually has $N$ Majorana
zero-modes associated with it, if we neglect interlayer hopping.
Consequently, a pair of vortices has $N$ qubits associated
with it. When two vortices are braided, their topological interactions,
as summarized in \ref{eqn:statistics} occur
independently in each layer.
In other words, the Majorana modes in layer $\alpha$ only affect
other Majorana modes in layer $\alpha$, not those in layer $\beta$.
Interlayer hopping causes these fermionic
zero-modes to hybridize and form a band along
the $c$-axis. While these states are split in energy,
their topological protection is unaffected. 

Another important difference is that vortices are potentially
heavy objects, so it is less clear that one can
observe the coherent quantum motion of vortices.
However, vortices are expected to become light as the Mott transition
is approached. While the normal state of Sr$_2$RuO$_4$
appears to be a Fermi liquid at low temperatures, transport
at temperatures above $30$ K is anomalous \cite{Mackenzie-Review}.
If this is an indication of proximity to the Mott transition, then it
could imply that vortices are light. In general, we expect
vortex motion to be coherent in the absence of appreciable
quasiparticle excitations. Having very low temperature would
help prevent decoherence by quasiparticles and by phonons.

Keeping all of these caveats in mind, we turn
to experiments which could observe the topological
properties of vortices in Sr$_2$RuO$_4$
thereby, in principle, taking the first steps towards topologically protected
qubits. Our goal is to observe the
two topologically-degenerate states of a vortex.

Recently, Stone and Chung \cite{Stone05}
emphasized that when two vortices are brought
together, the difference between the two states of
a qubit is now manifest: there either is or isn't a neutral fermion
in the core of the combined vortex. There is an energy
cost in the latter case. Since we expect neither an
excess of charge nor spin in the vortex core, we
can look for subtle differences in the charge distribution
or try to observe the energy difference.
The idea is the following. While the total charge is the
same whether or not the neutral fermion is present,
the local charge distribution will be different in the
two cases. Thus, if one had spatial resolution
smaller than the size of a vortex core $\sim 660$ \AA,
one could try to look with an STM for the particular charge
density profiles associated with the presence
or absence of a neutral fermion.
In principle, gravitational effects can be used to
detect this energy difference, but the required sensitivity
may be very difficult to achieve.

\begin{figure}[tbh]
\includegraphics[width=3.25in]{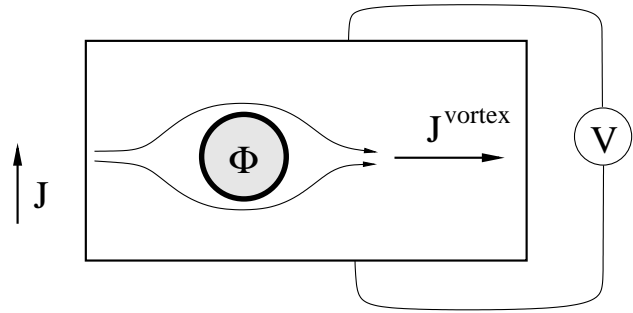}
\caption{In the experimental geometry depicted above,
a thin film of Sr$_2$RuO$_4$ has a central hole with a flux $\Phi$ through it.
When an electrical current flows in the $y$-direction,
depinned vortices flow in the $x$-direction. This generates
a voltage drop in the $x$ direction. If the motion of a vortex
is coherent, then there will be interference between
trajectories which go around either
side of the hole. The nature of
this interference will depending strongly
on whether the number of half-quantum vortices in
the hole is odd or even.}
\label{fig:flux-voltage}
\end{figure}

An alternate strategy is to observe the braiding statistics
of vortices through their effect on quantum interference.
Consider the experimental setup in Figure \ref{fig:flux-voltage}.
We assume that a thin film has a hole bored in it
so that it has an annular topology. Flux $\Phi$ is threaded
through the hole, so that there are $N_v$ vortices in the hole.
A magnetic field $H>H_{c1}$ is applied to the rest of the film so
that vortices penetrate the rest of the film as well.
A current $J$ is driven in the $y$-direction.
Vortex motion in the $x$-direction generates a voltage
drop in the $y$-direction, as in (\ref{eqn:Josephson}).
If the motion of a vortex is coherent,
then the two
trajectories in the figure will contribute to the
resistivity according to \cite{Fradkin98}:
\begin{equation}
R_{yy} \propto {\left|{t_1}\right|^2} + {\left|{t_2}\right|^2}
+ 2 \text{Re}\left\{ {t_1}{t_2^*}
\left\langle \Psi\left| B \right| \Psi \right\rangle \right\}
\end{equation}
where ${t_1}$, ${t_2}$ are the amplitudes for the
two trajectories, $B$ is the braiding operator for
a vortex to go around the vortices in the hole,
and $| \Psi \rangle$ is the combined state of the moving vortex
and the vortices in the hole.
When the number of vortices in the hole is
even, $\left\langle \Psi\left| B \right| \Psi \right\rangle $ is a phase,
$e^{i\theta}$. However, when the number of vortices in
the hole is odd, $\left\langle\Psi\left| B \right| \Psi \right\rangle = 0$
\cite{Stern05,Bonderson05}. If we can vary $t_1$ and $t_2$,
then we can distinguish these two cases.
It is not clear how to do this, but there are several possibilities.
(1) If we gate the Sr$_2$RuO$_4$ thin film along one
of the possible trajectories and apply
a gate voltage, this will suppress the electron density
along this trajectory. This, in turn, will change the
amplitude for this trajectory. (2) If pressure is applied to
the part of the film where one of the trajectories is located,
the lattice constant will be different there, presumably
translating into a different amplitude for this trajectory.
(3) Applying a temperature gradient will cause the superfluid density
to be different along the two different trajectories which, again,
would lead to a variation in their respective amplitudes. However,
there is a caveat to this final option: raising the temperature may
be dangerous since we want vortex motion to be coherent.
One could also try to observe interference in the Nernst effect
in the geometry of Figure \ref{fig:flux-voltage}, but this is difficult
experimentally.

\paragraph{Conclusion.}
In summary, we propose an experimentally feasible technique
for stabilizing half-quantum vortices in Sr$_2$RuO$_4$, assuming the system
to be a ${p_x}+i{p_y}$ superconductor. If stabilized, such half-quantum vortices should exhibit non-Abelian braiding statistics and, potentially, lead to topological quantum computation.

\begin{acknowledgements}
We have been supported by the ARO-ARDA under Grant
No.~W911NF-04-1-0236. S.~D.~S. and S.~T. have
also been supported by LPS-NSA.
C.~N. has also been supported by the NSF
under Grant No.~DMR-0411800.
\end{acknowledgements}


\end{document}